%% file: main.tex
\documentclass[sigconf]{acmart}
\copyrightyear{2024}
\acmYear{2024}
\setcopyright{rightsretained}
\acmConference[GLSVLSI '24]{Great Lakes Symposium on VLSI 2024}{June 12--14, 2024}{Clearwater, FL, USA}
\acmBooktitle{Great Lakes Symposium on VLSI 2024 (GLSVLSI '24), June 12--14, 2024, Clearwater, FL, USA}
\acmDOI{10.1145/3649476.3658714}
\acmISBN{979-8-4007-0605-9/24/06}
\acmSubmissionID{123-A56-BU3}

\author{Xinshi Zang}
\affiliation{%
  \institution{The Chinese University of Hong Kong}
  \country{}
}
\email{xszang@cse.cuhk.edu.hk}

\author{Wenhao Lin}
\affiliation{%
  \institution{The Chinese University of Hong Kong}
  \country{}
}
\email{whlin23@cse.cuhk.edu.hk}

\author{Shiju Lin}
\affiliation{%
  \institution{The Chinese University of Hong Kong}
  \country{}
}
\email{sjlin@cse.cuhk.edu.hk}

\author{Jinwei Liu}
\affiliation{%
  \institution{The Chinese University of Hong Kong}
  \country{}
}
\email{jwliu@cse.cuhk.edu.hk}

\author{Evangeline F.Y. Young}
\affiliation{%
  \institution{The Chinese University of Hong Kong}
  \country{}
}
\email{fyyoung@cse.cuhk.edu.hk}

\usepackage{subfig}
\usepackage{amsmath}
\usepackage{amssymb}
\usepackage{amsfonts}
\usepackage{algorithmic}
\usepackage{graphicx}
\usepackage{textcomp}
\usepackage{xcolor}
\usepackage{caption}
\usepackage{subcaption}
\usepackage{multirow}
\usepackage{makecell}
\usepackage{booktabs} % To thicken table lines
\def\BibTeX{{\rm B\kern-.05em{\sc i\kern-.025em b}\kern-.08em
    T\kern-.1667em\lower.7ex\hbox{E}\kern-.125emX}}
\usepackage{graphics}
\usepackage[linesnumbered,ruled,vlined]{algorithm2e}
\SetKwInput{KwInput}{Input}                % Set the Input
\SetKwInput{KwOutput}{Output}  
\newcommand\blfootnote[1]{%
  \begingroup
  \renewcommand\thefootnote{}\footnote{#1}%
  \addtocounter{footnote}{-1}%
  \endgroup
}

\begin{document}

\newcommand{\et}{et al.}
\newcommand\todo[1]{\textcolor{red}{#1}}
\newcommand{\ours}{CPP}
\newcommand{\ourss}{CPP }
\newcommand{\firb}{WLM}
\newcommand{\firbs}{WLM }
\newcommand{\secb}{EaSyOpt}
\newcommand{\secbs}{EaSyOpt }
\newcommand{\thib}{Circuitformer}
\newcommand{\thibs}{Circuitformer }

\title{
An Open-Source Fast Parallel Routing Approach for Commercial FPGAs}

\ifx\Format\ACMFormat %is true
\input{sections/abstract}

\maketitle
\else 
\maketitle
\input{sections/abstract}
\fi

\input{sections/introduction}
\input{sections/preliminaries}
\input{sections/methodolgy}

\input{sections/experiment}

\input{sections/conclusion}

\bibliographystyle{ACM-Reference-Format}
\bibliography{main}

\end{document}

%% file: sections/abstract.tex
\begin{abstract}
In the face of escalating complexity and size of contemporary FPGAs and circuits, routing emerges as a pivotal and time-intensive phase in FPGA compilation flows. In response to this challenge, we present an open-source parallel routing methodology designed to expedite routing procedures for commercial FPGAs. Our approach introduces a novel recursive partitioning ternary tree to augment the parallelism of multi-net routing. Additionally, we propose a hybrid updating strategy for congestion coefficients within the routing cost function to accelerate congestion resolution in negotiation-based routing algorithms. Evaluation on public benchmarks from the FPGA24 routing contest demonstrates the efficacy of our parallel router. It achieves a 2$\times$ speedup compared to the academic serial router RWRoute. Furthermore, when compared to the industry-standard tool Vivado, our approach not only delivers a 2$\times$ acceleration but also yields a notable 31\% enhancement in critical-path wirelength.
\end{abstract}

% <concept_id>10010583.10010682.10010697.10010704</concept_id>
% <concept_desc>Hardware~Wire routing</concept_desc>

% \ccsdesc[500]{Hardware~Wire routing}

% \keywords{FPGA routing, parallel routing}

%% file: sections/introduction.tex
\section{Introduction}
\blfootnote{The work described in this paper was partially supported by a grant
from the Research Grants Council of the Hong Kong Special Administrative
Region, China (Project No. CUHK14218422).}

Routing constitutes a pivotal phase in the compilation flow of Field-Programmable Gate Arrays (FPGAs), entailing the assignment of each net to wire segments on FPGAs to ensure connectivity without overlap. With the escalating complexity of modern FPGAs and circuits, routing has evolved into an exceedingly challenging and time-consuming process~\cite{chen2017fpga}. Accelerating routing procedures holds profound significance across various FPGA application domains, such as ASIC emulation.

Numerous existing works have endeavored to expedite FPGA routing runtime through parallel routing techniques employing multi-threads and Graphics Processing Units (GPUs)~\cite{gort2011accelerating, shen2017corolla, hoo2018paradro, moctar2018deterministic, wang2019parra, shen2020combining, zhou2020accelerating, murray2020vtr}. Multi-net parallelization is central to these efforts~\cite{gort2011accelerating, hoo2018paradro, wang2019parra, shen2020combining, zhou2020accelerating, murray2020vtr}, wherein different recursive spatial partitioning methods are employed to divide nets into geographically independent sets, facilitating parallel routing. Additionally, some studies have explored single-net parallelization by parallelizing maze expansion or leveraging GPU-accelerated Bellman-Ford algorithms~\cite{moctar2018deterministic, shen2017corolla}. While significant acceleration has been achieved, the majority of existing parallel FPGA routing works are not open source, hindering community development. Notably, VTR~\cite{murray2020vtr} is a popular open-source FPGA CAD tool; however, it primarily targets academic hypothetical architectures, rendering it impractical for commercial FPGAs. 

Addressing these limitations, we propose an open-source parallel FPGA routing approach tailored for commercial devices. We leverage the open-source negotiation-based serial router, RWRoute \cite{zhou2021rwroute}, which is contained in the RapidWright~\cite{lavin2018rapidwright} framework and is compatible with commercial FPGAs. Different from the binary partitioning tree used in existing works~\cite{gort2011accelerating, hoo2018paradro, zhou2020accelerating}, we propose a novel recursive partitioning ternary tree (RPTT) to schedule the parallel routing of multiple nets. We iteratively construct a full ternary tree for each set of nets within an FPGA region until further partitioning is unfeasible. By enabling independent routing of nets in the left and right RPTT sub-trees, our RPTT significantly enhances the parallelization of multi-net routing. Furthermore, we propose a Hybrid Updating Strategy (HUS), incorporating present-centric and historical-centric updating, for congestion coefficient adjustments within the routing cost function. This hybrid strategy effectively balances pathfinding time and congestion resolution, particularly in congested designs.
The specific contributions of this work are summarized as follows:
\begin{itemize}
    \item We propose an open-source\footnote{The codes are available at https://github.com/xszang/parallel-routing.} parallel FPGA routing approach to expedite routing for commercial FPGAs.
    \item We introduce a Recursive Partitioning Ternary Tree (RPTT) to enhance multi-net parallelization and a Hybrid Updating Strategy (HUS) to accelerate congestion resolution.
    \item On the FPGA24 contest public benchmarks~\cite{contest}, our parallel router outperforms the serial router RWRoute by achieving approximately twice the speed. Furthermore, our router surpasses the commercial tool Vivado, exhibiting a 31\% improvement in critical-path wirelength and doubling the speed. 
\end{itemize}

The remainder of this paper is organized as follows: Section~\ref{sec:preliminary} presents preliminaries about the commercial UltraScale+ FPGA architecture. Section~\ref{sec:method} discusses our partitioning-based parallel routing framework with the hybrid updating strategy. Section~\ref{sec:exp} provides detailed experimental results and ablation studies. Finally, Section~\ref{sec:conclusion} concludes the work.

%% file: sections/preliminaries.tex
\section{Preliminaries}
\label{sec:preliminary}
In this section, we provide an overview of the target FPGA architecture, the routing resource graph, and discuss fundamental serial FPGA routing algorithms.

\subsection{Ultrascale+ FPGA Architecture}
\begin{figure}[b]
    \centering
\includegraphics[width=0.9\columnwidth,page=1]{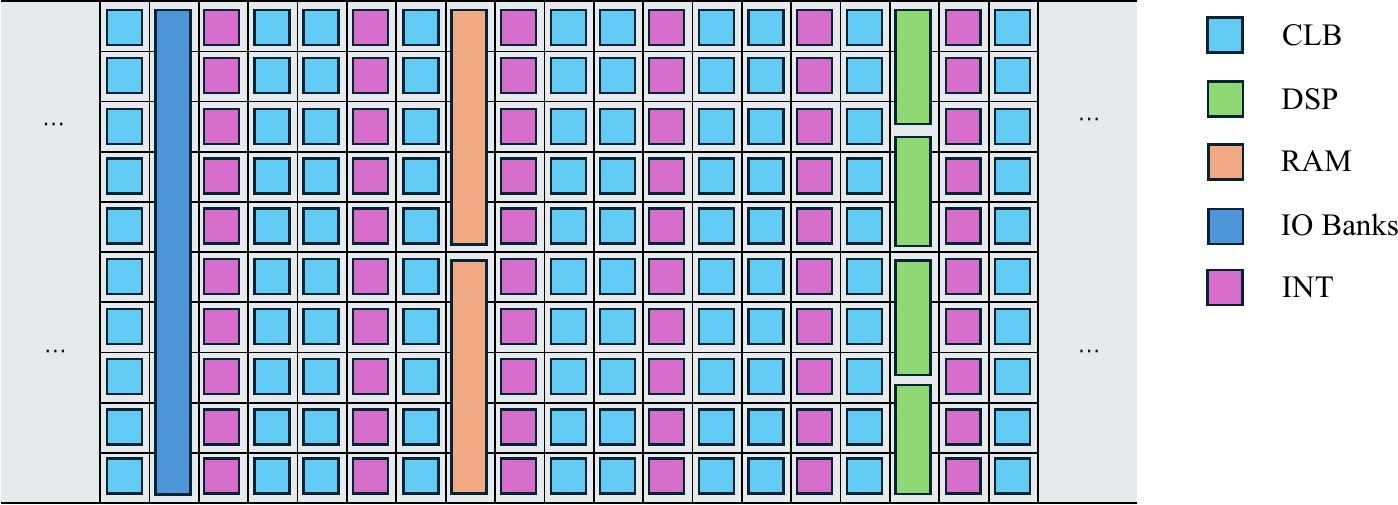}
    \caption{UltraScale+ FPGA with columnar resources}
    \label{fig:layout}
\end{figure}

We focus on Xilinx Virtex UltraScale+ FPGAs in this study, characterized by a columnar architecture as depicted in Fig. \ref{fig:layout}. This architecture comprises configurable logic blocks (CLB), block RAM (BRAM), DSP units, IO interfaces, and interconnect (INT) blocks. The INT blocks are interspersed among other functional blocks, forming a programmable interconnection network that facilitates connections between various functional blocks. The wires interconnecting INT blocks are fixed routing resources and can span 1, 2, 4, or 12 INT blocks. Notably, the INT block is configurable, allowing for diverse wire connectivities. Fig.~\ref{fig:box2graph} illustrates a simplified INT block example, with dashed lines indicating programmable interconnect points (PIPs).

\subsection{Routing Resource Graph}
\begin{figure}
    \centering
    \subfloat[Interconnect architecture]{\includegraphics[width=0.44\linewidth,page=2]{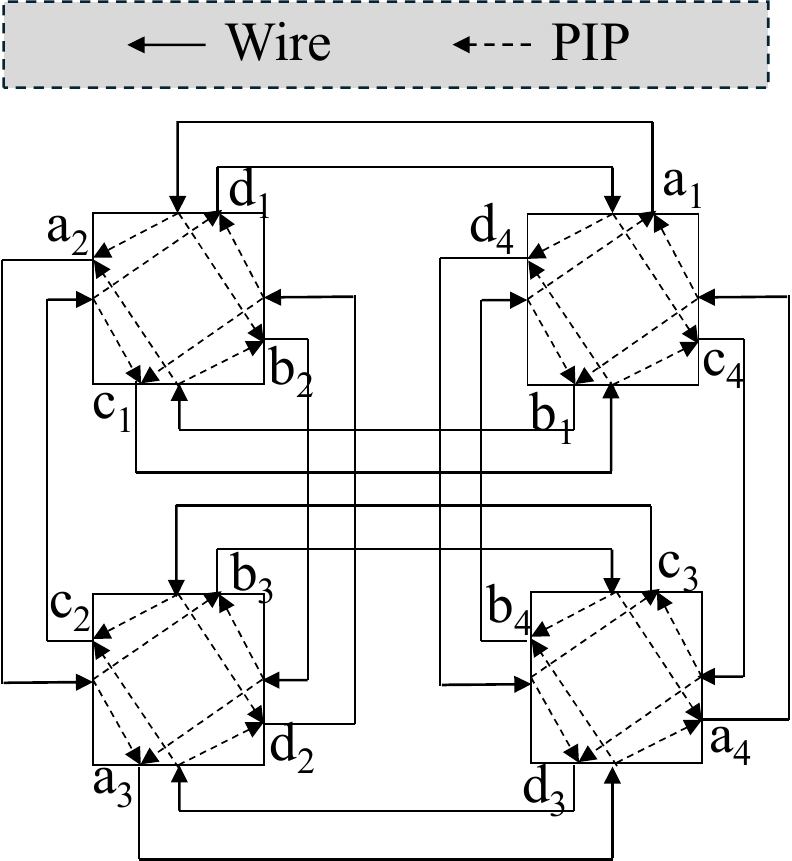}}
    \hspace{.05\columnwidth}
    \subfloat[Routing graph]{\includegraphics[width=0.44\linewidth,page=3]{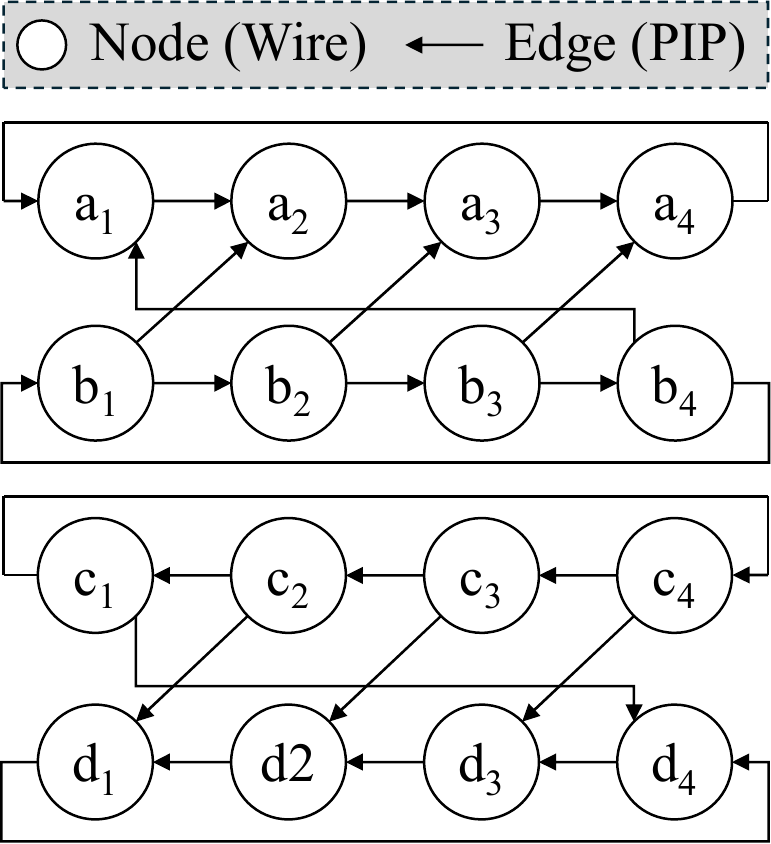}}
    \caption{Construction of a routing resource graph}
    \label{fig:box2graph}
\end{figure}

Routing resources encompass wires and PIPs on the FPGA device, typically represented as a directed graph $G=(V,E)$. In the routing resource graph (RRG), nodes correspond to wires, and edges represent PIPs. Fig.~\ref{fig:box2graph} illustrates the construction of a routing resource graph. Each RRG node's length is proportional to the number of INT blocks spanned by the corresponding wire. Compared to academic hypothetical architectures, the UltraScale+ FPGA exhibits an extensive routing resource graph, comprising over 28 million nodes and 125 million edges, imposing significant complexity on routing tasks.

After placement, the signal nets that have no direct wires connecting them need to be routed on the RRG. The source and sink pins of nets are first mapped to RRG nodes. The routing problem is then transformed into a path-searching task on the RRG. For each net, the source and sink pins need to be connected by using as few RRG nodes as possible. Each RRG node can only accommodate at most one net, or in other words, the capacity of a RRG node for routing is 1. The resource overflows, meaning that multiple nets pass through one RRG node, are prohibited in FPGA routing.

\subsection{FPGA routing algorithm}
The negotiation-based routing algorithm, exemplified by PathFinder \cite{McMurchie1995PathFinderAN}, is pivotal and widely employed in FPGA routers. PathFinder employs a negotiation mechanism, defining the cost function of a given node $n$ as shown in Eq.~\eqref{eq:pathfinder_cost},
\begin{equation}
    f(n)=(b(n)+h(n))\times p(n)
    \label{eq:pathfinder_cost}
\end{equation}
where $b(n)$ denotes the base cost associated with the node's length, $h(n)$ represents accumulated historical congestion cost, and $p(n)$ signifies present congestion cost. By incrementally increasing congestion costs, the routing algorithm eventually finds legal routing solutions for all nets without congestion.

Existing routing algorithms can be categorized into net-based~\cite{wang2019parra, murray2020vtr, shen2020combining} and connection-based~\cite{hoo2018paradro, zhou2020accelerating, zhou2021rwroute} routers, depending on their treatment of multi-pin nets. Net-based algorithms route entire multi-pin nets as a whole, while the connection-based routers first split a multi-pin net into several two-pin nets (called connections) and then route each connection independently. To minimize net wirelength, connection-based routers encourage reuse of common RRG nodes among different connections from the same net. As the overlap between connections is typically smaller than that between nets, connection-based routing algorithms enable greater parallelism for parallel routing. Consequently, we adopt the connection-based approach in this work to develop the parallel algorithm.

%% file: sections/methodolgy.tex
\begin{figure}[t]
	\centering
	\includegraphics[width=0.9\linewidth, page=4]{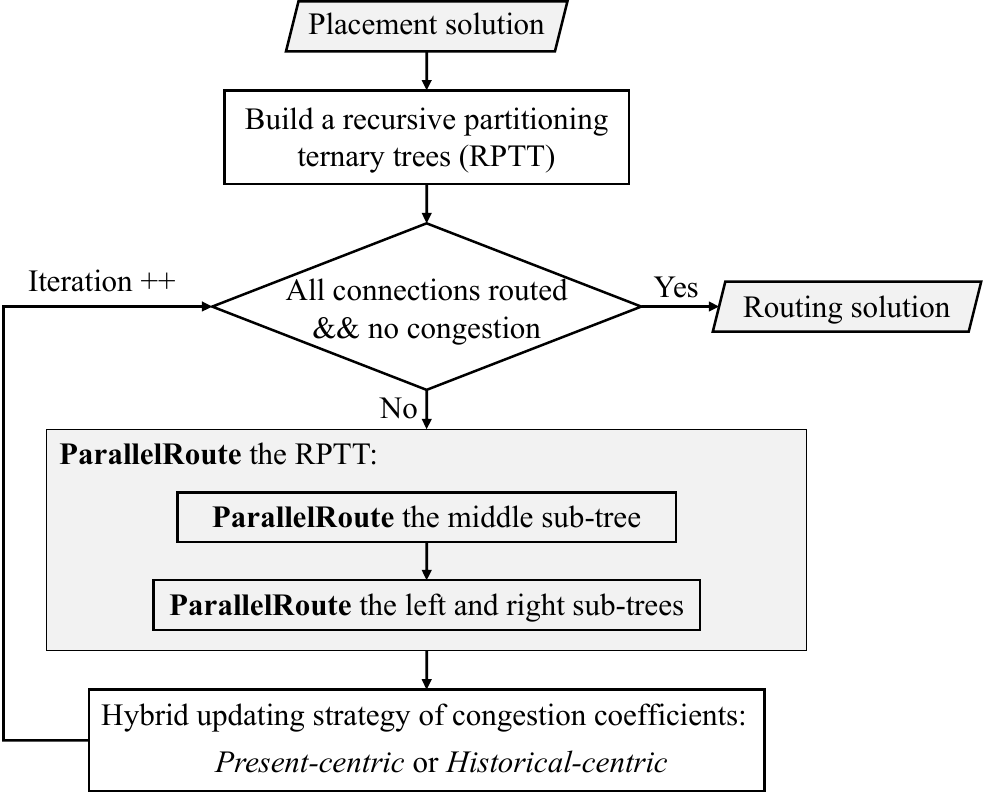}
	\caption{The framework of our parallel router}
	\label{fig:framework}
\end{figure}
\section{Methodology}
\label{sec:method}
In this section, we will describe the framework of our parallel router in detail, as depicted in Fig.~\ref{fig:framework}. Given the placement solution, we first build a recursive partitioning ternary tree (RPTT) to facilitate the scheduling of multi-net parallel routing. Subsequently, we iteratively perform parallel routing based on the RPTT and update the congestion coefficients until a valid and resource overflow-free routing solution is found. Additionally, we employ a hybrid updating strategy (HUS), containing present-centric and historical-centric updating, for congestion coefficients at the end of each routing iteration to expedite congestion resolution.

\subsection{Recursive Partitioning Ternary Tree (RPTT)}
Partitioning-based parallel routing is widely adopted by existing parallel routing works~\cite{gort2011accelerating,hoo2018paradro,shen2020combining}. The bi-partitioning approach is first applied to recursively split the FPGA board into two disjoint regions with interleaved horizontal and vertical cutlines. A full binary partitioning tree is then built where each sub-tree represents the nets fully contained in a region and the root node of a sub-tree represents the nets crossing the cutline. With the partitioning tree, the routing of all nets is scheduled by first routing the root node and then parallelly routing the two sub-trees. 

\begin{algorithm}[t]
	\caption{RPTT-based Parallel Routing} 
	\label{alg:routing}
	\KwInput{
		Connections $C$
	}
	\SetKwFunction{FMain}{Main}
	\SetKwFunction{FParRoute}{ParallelRoute}
	\SetKwFunction{FSeqRoute}{BlockedRoute}
	\SetKwFunction{FCut}{BalanceCut}
	\SetKwFunction{FBuildTree}{BuildTree}
	Initialize an empty ternary tree $T$ \\
	\FBuildTree{$C$, $T.root$}\\
	Initialize an empty ThreadPool $pool$ \\
	\FParRoute{$root$, $pool$}\label{alg:main_entry}\\
		Finish all works in $pool$\\
	\SetKwFunction{FMain}{Main}
	\SetKwFunction{FCut}{BalanceCut}
	\SetKwFunction{FCutX}{BalanceCutX}
	\SetKwFunction{FBuildTree}{BuildTree}
	\SetKwProg{Fn}{Function}{:}{}
	\Fn{\FBuildTree{Connections $C$, Tree root}\label{alg:build_tree}}{
		$success, C_l, C_m, C_r \gets$ \FCut{$C$}  \label{alg:call_cut}\\ 
		\If{$success$}{
			$root.left \gets$ \FBuildTree{$C_l$, $root.left$} \\
			$root.mid \gets$ \FBuildTree{$C_m$, $root.mid$} \\
			$root.right \gets$ \FBuildTree{$C_r$, $root.right$} \\
		}
		\Else{
			Set $root.left$, $root.mid$, and $root.right$ as $null$ \\
		}
		$root.connections \gets C$ \\
		\KwRet $root$\\
	}
	\SetKwProg{Fn}{Function}{:}{}
	\Fn{\FParRoute(RPTT $root$, ThreadPool $pool$)}{\label{alg:para_route}
		\If{$root$ has sub-trees}{
			\FSeqRoute{$root.mid$} \\
			$pool.add$(\FParRoute{$root.left$, $pool$}) \\
			$pool.add$(\FParRoute{$root.right$, $pool$})\\
		}\Else{
			Sequentially route connections in $root.connections$\\
		}
	}
    \Fn{\FSeqRoute(RPTT $root$)\label{alg:seq_route}}{
		Initialize an empty ThreadPool $pool$ \\
		\FParRoute{$root$, $pool$}\\
		Finish all works in $pool$
	}
\end{algorithm}

\begin{figure}[b]
	\centering
\includegraphics[width=0.90\linewidth, page=5]{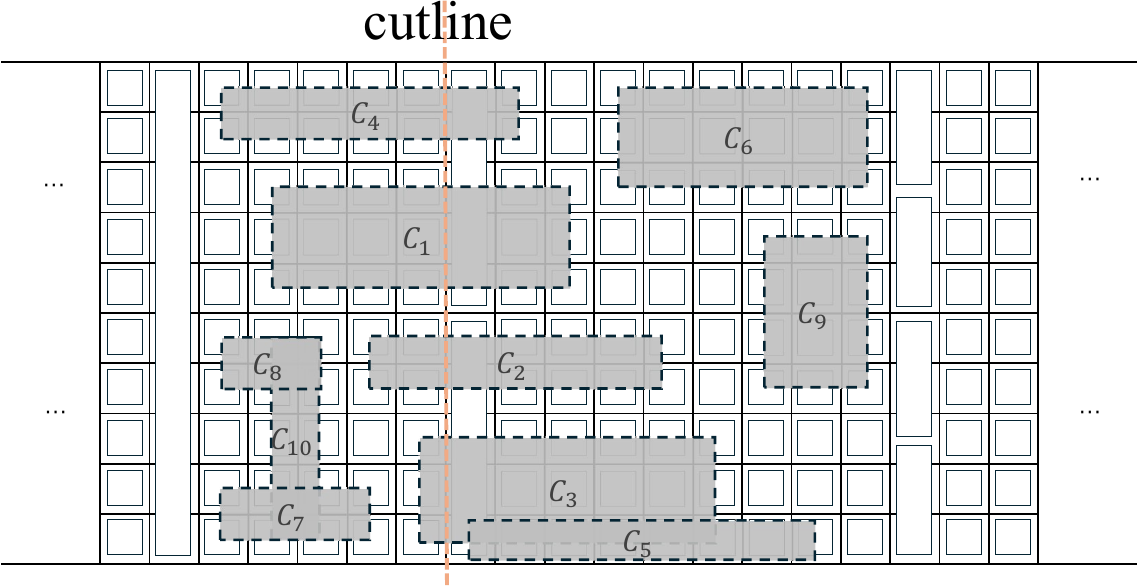}
	\caption{Ten connections with one cutline}
	\label{fig:cutline}
\end{figure}
\begin{figure}[t]
	\centering
\includegraphics[width=0.90\linewidth, page=6]{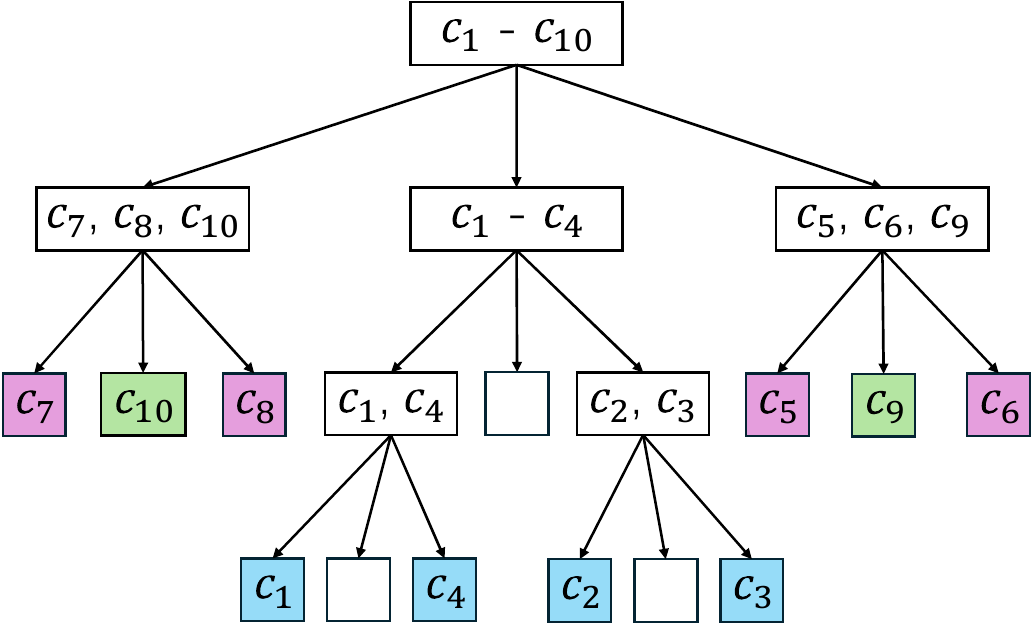}
	\caption{RPTT for the connections in Fig. 4}
	\label{fig:recursive_tree}
\end{figure}

To further speed up the routing of the root node and establish a universal data structure for scheduling, we propose a recursive partitioning ternary tree (RPTT) in this work, which is described in the function (line~\ref{alg:build_tree}) of Alg.~\ref{alg:routing}. The left and right sub-trees of RPTT are similar to those of the aforementioned binary partitioning tree but RPTT also constructs a middle sub-tree for the nets crossing the cutline. The detailed procedure to determine the cutline in line~\ref{alg:call_cut} of Alg.~\ref{alg:routing} will be discussed later. The connections on each non-leaf node in RPTT represent the union of connections in its sub-trees. To demonstrate the RPTT build process, one RPTT is provided in Fig.~\ref{fig:recursive_tree} for the connections distributed in Fig.~\ref{fig:cutline}.

After building the RPTT, we then perform a universal parallel routing procedure in line~\ref{alg:main_entry} of Alg.~\ref{alg:routing}. The {\tt ParallelRoute} is a recursive function that first performs the routing of the middle sub-tree in a blocked way and then submits two parallel jobs of the routing of the left and right sub-trees. The {\tt BlockedRoute} requires that all routing tasks submitted in this sub-trees need to be finished at the end of the function while the {\tt ParallelRoute} doesn't have this requirement. A unique task pool is maintained for each middle sub-tree to manage the routing jobs which are independent and invoked recursively on the sub-tree. With the scheduling method in Alg.~\ref{alg:routing}, the execution orders of different connections in Fig.~\ref{fig:recursive_tree} are marked with different colors. The four blue connections will be routed firstly in parallel and then the two green ones follow and the four pink ones are parallelly routed at last. 

To build an efficient partitioning tree for parallelism, it is of great importance to select a good cutline that splits the nets in balance. To this end, we propose a balance-driven algorithm inspired by ParaDRo~\cite{hoo2018paradro} to determine the optimal cutline for balanced partitioning. As described in Alg.~\ref{alg:get_cut}, we will try to do the partitioning on the x-axis and y-axis and adopt the one with a smaller difference in partition size. Whatever the axis is, two lists, including $size\_bebore$ and $size\_after$, are maintained to accumulate the connection numbers aside from a cutline. By doing so, the optimal cutline with the smallest imbalance in partition size can be found in a linear time. The recursive partitioning will be terminated in a region if no cutline can split the connections into two independent parts.

\begin{algorithm}[t]
	\caption{Balance-driven Cutline}
	\label{alg:get_cut}
	
	\SetKwFunction{FMain}{Main}
	\SetKwFunction{FCut}{BalanceCut}
	\SetKwFunction{FCutX}{BalanceCutX}
	\SetKwFunction{FCutY}{BalanceCutY}
	\SetKwFunction{FBuildTree}{BuildTree}
	\SetKwProg{Fn}{Function}{:}{}
	\Fn{\FCut(Connections $C$)}{
		$success^x, C_l^x, C_m^x, C_r^x, diff^x \gets$ \FCutX{$C$} \\
		$success^y, C_l^y, C_m^y, C_r^y, diff^y \gets$ \FCutY{$C$} \\
		\If{$diff^x < diff^y$}{
			\KwRet $success^x, C_l^x, C_m^x, C_r^x$
		}\Else{
			\KwRet $success^y, C_l^y, C_m^y, C_r^y$
		}
	}
	\Fn{\FCutX{Connections $C$}}{
		get the horizontal boundary $(x_{min}, x_{max})$ of $C$\\
		Initialize lists $size\_{before}$ and $size\_{after}$ with $0$\\
		\For{Connection $c \in C$}{
			\For{$i\in[c.x_{max}, x_{max}]$}{
				$size\_{before}[i] \gets size\_{before}[i] + 1$
			}
			\For{$i\in[x_{min}, c.x_{min})$}{
				$size\_{after}[i] \gets size\_{after}[i] + 1$
			}
		}
		$diff \gets inf$, $cutline \gets 0$ \\
		\For{$i \in [x_{min}, x_{max}]$}{
			$d \gets |size\_{after}[i] - size\_{before}[i]|$ \\
			\If{$d < diff$}{
				$diff \gets d$, $cutline \gets i$ \\   
			}
		}
		
		$C_l \gets \{c\in C| c.x_{max} \leq cutline\}$\\
		$C_r \gets \{c\in C| c.x_{min} > cutline\}$\\
		$C_m \gets \{c \in C| c \notin C_l, C \notin C_r\}$\\
		$success \gets true$\\
		\If{$C_l$ is empty or $C_r$ is empty}{
			$success \gets false$\\
		}
		\KwRet $sucess, C_l, C_m, C_r, diff$
	}
	\Fn{\FCutY{Connections $C$}}{
		\tcc{Similar with \FCutX, skipped here}
	}
\end{algorithm}

\subsection{Hybrid Updating Strategy (HUS)}
In traditional path-finding algorithms, the $A^*$ searching algorithm is usually first applied to find the short path between the source node and the sink node for each net. The path length during $A^*$ searching is related to both wirelength and the congestion cost. By gradually increasing the congestion penalty, the routing algorithms can converge to a congestion-free solution with optimized wirelength. However, as pointed out in the enhanced PathFinder algorithm~\cite{zha_revisiting_2022}, the updating strategy of the congestion penalty has a great influence on both quality and runtime. 
In this work, a hybrid updating strategy (HUS) is proposed to speed up the congestion fixing during the negotiation-based routing algorithm. 

Before discussing the HUS, we first introduce the routing cost function of RRG nodes in this work. As defined in Eq~\eqref{eq:cost}, the node cost, $f(n)$, is composed of three parts. The first term is the upstream path length, represented by $c_{prev}(n)$, which is the accumulated cost of all nodes in the partial path from the source to the node $n$. The second $c_{est}(n)$ denotes the estimation of the cost of reaching the sink from the node $n$, which is estimated by the Mahattan distance. 
\begin{equation}
    \label{eq:cost}
    f(n)=c_{prev}(n) + c_{est}(n) + c(n) 
\end{equation}
The third item, $c(n)$ in Eq.~\eqref{eq:cost}, defines the cost of using the node $n$, which is further formulated in Eq.~\eqref{eq:base_cost}. 
$c(n)$ depends on the wirelength-related base cost $b(n)$, the present congestion cost $p(n)$ and the historically accumulated congestion cost $h(n)$. 
$share(n)$ represents the number of connections from the same net that are using $n$, encouraging connections to reuse the existing path of the net. 
\begin{equation}
    \label{eq:base_cost}
    c(n)=\dfrac{b(n)\cdot h(n)\cdot p(n)}{1+share(n)}
\end{equation}
To resolve routing congestions, $h(n)$ and $p(n)$ are increased with different updating strategies during the entire routing process.
The updating strategies in RWRoute are defined in Eq.~\eqref{eq:updating_h} and~\eqref{eq:updating_p}, where $occ(n)$ is the number of nets using RRG node $n$ and $i$ represents the number of iterations of routing all nets. $h_f$ and $p_f$ are const congestion coefficients (1 and 2 respectively) in RWRoute. The historical cost $h(n)$ is linear with $h_f$ while the present cost $p(n)$ is exponential to $p_f$.
\begin{align}
h^i(n)&=\begin{cases}
    % 1&\text{, if }i=1\\
    h^{i-1}(n)+h_f\cdot(occ(n)-1)&\text{, if }occ(n)>1\\h^{i-1}(n)&\text{, otherwise}\end{cases} \label{eq:updating_h}\\
p^i(n)&=1+p_0\cdot{(p_f)}^{i-1}\cdot occ(n)\label{eq:updating_p}
    % p_f &= p_0\cdot{(p_m)}^{i-1} \\
\end{align}

\begin{table}[b]
\caption{Statistics of FPGA24 public benchmarks}
\label{tab:benchmark}
\resizebox{1.0\linewidth}{!}{
\begin{tabular}{l|cc|cccc}
\hline
Benchmark & Nets (k) & Connections (k) & LUTs (k) & FFs (k) & DSPs & BRAMs \\ \hline
logicnets\_jscl & 28 & 180 & 31 & 2 & 0 & 0 \\
boom\_med\_pb & 54 & 221 & 36 & 17 & 24 & 142 \\
vtr\_mcml & 71 & 225 & 43 & 15 & 105 & 142 \\
rosetta\_fd & 77 & 230 & 46 & 39 & 72 & 62 \\
corundum\_25g & 166 & 495 & 73 & 96 & 0 & 221 \\
finn\_radioml & 110 & 405 & 74 & 46 & 0 & 25 \\
vtr\_lu64peeng & 143 & 537 & 90 & 36 & 128 & 303 \\
corescore\_500 & 179 & 590 & 96 & 116 & 0 & 250 \\
corescore\_500\_pb & 175 & 597 & 96 & 116 & 0 & 250 \\
mlcad\_d181\_lefttwo3rds & 361 & 916 & 155 & 203 & 1344 & 405 \\
koios\_dla\_like\_large & 509 & 912 & 189 & 362 & 2209 & 192 \\
boom\_soc & 274 & 1374 & 227 & 98 & 61 & 161 \\
ispd16\_example2 & 449 & 1455 & 289 & 234 & 200 & 384 \\ \hline
UltraScale+ xcvu3p & - & - & 394 & 788 & 2280 & 720 \\ \hline
\end{tabular}
}
\end{table}

\begin{table*}[t]
\caption{Overall performance. All metrics are the smaller the better.}
\label{tab:exp_main_result}
\resizebox{1.0\linewidth}{!}{
\begin{tabular}{l|ccc|ccc|ccc}
\hline
\multirow{2}{*}{Benchmark} & \multicolumn{3}{c|}{Vivado} & \multicolumn{3}{c|}{RWRoute} & \multicolumn{3}{c}{Ours} \\
 & Runtime (s) & Wirelength & Score & Runtime (s) & Wirelength & Score & Runtime (s) & Wirelength & Score \\ \hline
logicnets\_jscl & 78.33 & 310 & 101.50 & 52.03 & \textbf{226} & 69.43 & \textbf{35.26} & 234 & \textbf{55.13} \\
boom\_med\_pb & \textbf{139.33} & 823 & \textbf{207.70} & 230.88 & 969 & 304.69 & 144.50 & \textbf{806} & 210.65 \\
vtr\_mcml & 490.33 & 666 & 507.90 & 243.13 & 594 & 278.22 & \textbf{94.29} & \textbf{584} & \textbf{143.26} \\
rosetta\_fd & 147.67 & 888 & 221.70 & 161.30 & 839 & 229.07 & \textbf{125.32} & \textbf{804} & \textbf{193.19} \\
corundum\_25g & - & - & - & 249.61 & \textbf{396} & 264.25 & \textbf{131.11} & 500 & \textbf{168.00} \\
finn\_radioml & 154.67 & 338 & 173.00 & 119.88 & 277 & 135.59 & \textbf{63.29} & \textbf{251} & \textbf{82.06} \\
vtr\_lu64peeng & 218.67 & 1728 & 369.60 & 226.57 & 1412 & 345.12 & \textbf{114.12} & \textbf{1333} & \textbf{236.01} \\
corescore\_500 & 188.33 & 751 & 244.60 & 158.84 & 680 & 210.96 & \textbf{73.03} & \textbf{668} & \textbf{132.52} \\
corescore\_500\_pb & 226.67 & 861 & 290.10 & 278.30 & \textbf{687} & 319.17 & \textbf{138.63} & 739 & \textbf{198.67} \\
mlcad\_d181\_lefttwo3rds & \textbf{407.67} & 1159 & 482.80 & 1,779.59 & 809 & 1,682.53 & 409.81 & \textbf{771} & \textbf{445.93} \\
koios\_dla\_like\_large & 542.33 & 927 & 580.80 & 392.07 & 548 & 407.67 & \textbf{181.47} & \textbf{520} & \textbf{215.33} \\
boom\_soc & 711.00 & 2235 & 863.40 & 1,292.74 & 1698 & 1,333.26 & \textbf{635.33} & \textbf{1673} & \textbf{739.10} \\
ispd16\_example2 & 385.00 & 1481 & 494.60 & 584.94 & 1114 & 637.85 & \textbf{314.65} & \textbf{939} & \textbf{377.09} \\ \hline
Avg. Ratio & 2.04 & 1.31 & 1.73 & 2.10 & 1.03 & 1.76 & 1.00 & 1.00 & 1.00 \\ \hline
\end{tabular}
}\\ \footnotesize{*Vivado fails to route the corundum\_25g due to the failure in the DRC during the routing. \hspace{8.2cm}}
\end{table*}

The enhanced PathFinder \cite{zha_revisiting_2022} found that the quick growing speed of $p(n)$ will result in the net-order dependence issue and the original PathFinder algorithm can be improved by updating the historical congestion cost $h(n)$ alone. Hence, in the enhanced PathFinder, $p_f$ and $h_f$ are set to be 1 and 5 respectively. However, this setting will cause much more routing iterations to fix all routing conflicts. The maximum number of iterations in the enhanced PathFinder is set to 1500 to ensure successful routing for congested designs, which is much bigger than 50, the default setting of VTR~\cite{murray2020vtr}.

Different from~\cite{zha_revisiting_2022}, we propose a hybrid updating strategy (HUS) for the historical and present coefficients to reduce the runtime of fixing congestion while improving the routing wirelength for congested designs. 
In this work, we regard a circuit to route as a congested design if the ratio of the number of congested RRG nodes to the number of connections after the first iteration is larger than a threshold. In HUS, we first perform the present-centric updating by keeping the original fast updating strategy for the present coefficient to finish the routing of those uncongested areas. After a few iterations, most of the remaining connections to route are located in congested areas. We then apply the second historical-centric updating for the historical coefficient by decreasing the $p_f$ from 2 to $\alpha$ and increasing the $h_f$ from 1 to $\beta$. In the work, $\alpha$ and $\beta$ are set to 1.1 and 2 respectively. With the historical-centric updating, the negative impact of connection routing orders can be reduced, decreasing the number of iterations to fix the routing congestions.

%% file: sections/experiment.tex
\section{Experimental Results}
\label{sec:exp}

In this section, we will analyze the experimental results of different methods including Vivado and RWRoute~\cite{zhou2021rwroute}. We followed the setting of FPGA24 routing contest~\cite{contest} and conducted experiments in a high-performance computing server equipped with a 2.90GHz CPU and 768GB memory. Our approach was implemented in Java and ran with 16 threads by default. Vivado v2021.1 was compared with the command "{\tt route\_design -no\_timing\_driven -preserve}"\footnote{The FPGA24 contest focuses on runtime-first and non-timing-driven routing and requires the existing net routes to be preserved.} to conduct the routing task. The evaluation metrics of the FPGA24 contest include the runtime and critical wirelength. The overall score is equal to 0.9 * runtime + 0.1 * critical-path wirelength.
We ran all experiments three times and showed the average of the results to reduce the randomness effect.

\subsection{Benchmarks}
The statistics of FPGA24 public benchmarks are summarized in Tab.~\ref{tab:benchmark}. In FPGA24 contest, these circuits are obtained from different public benchmark suites and are then synthesized, placed, and routed on the target FPGA by using Vivado. The routing solutions of all signal nets are removed for the contest task. The benchmarks use the open-source FPGA Interchange Format (FPGAIF). The nets in Tab.~\ref{tab:benchmark} include all signal nets to be routed and the connections represent the 
corresponding two-pin sub-nets to be routed.

\subsection{Overall Performance}
The overall results of different methods are presented in  Tab.~\ref{tab:exp_main_result}. Compared with Vivado, RWRoute can significantly reduce the wirelength but incur considerable time overhead in some circuits, like mlcad\_d181\_lefttwo3rds and boom\_soc. Compared with both Vivado and RWRoute, our router can not only run two times faster on average but also further improve the wirelength in most cases, demonstrating the effectiveness of our proposed parallel framework. In the following, we will conduct two ablation studies to discuss the contributions of different techniques in our proposed method. 

\subsection{Ablation Studies}
Firstly, we conduct an ablation study on the recursive partitioning ternary tree (RPTT) in our framework by replacing the RPTT with the single recursive partitioning tree in ParaDRo~\cite{hoo2018paradro}. The comparison results, shown in Tab.~\ref{tab:exp_ablation_tree_enhance}, reveal that the RPTT can reduce the runtime by 14\% without obvious wirelength degradations. 

\begin{figure}[t]
    \centering
    \includegraphics[width=0.95\linewidth]{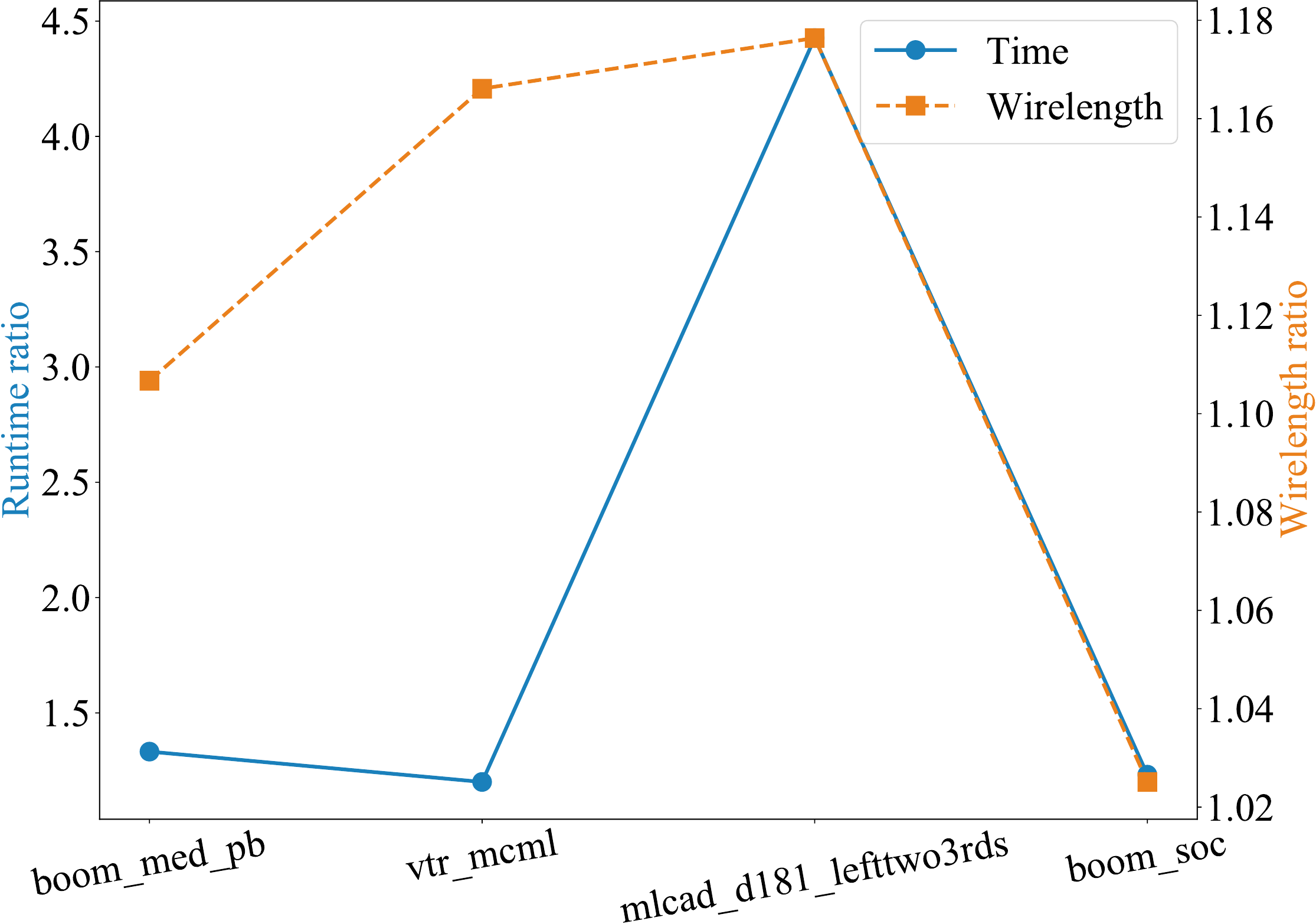}
    \caption{The comparison between Ours w.o HUS and Ours. The ratios larger than 1 represent degradation.}
    \label{fig:exp_ablation}
\end{figure}

Secondly, we study the effect of the hybrid updating strategy (HUS) for congestion coefficients. We disable the HUS and apply the default updating strategy in RWRoute. The results on the four congested designs, depicted in Fig.~\ref{fig:exp_ablation}, show that our HUS can both improve the runtime and the wirelength for congested designs. In particular, the runtime of mlcad\_d181\_lefttwo3rds is accelerated by around 4.5 times, and the wirelengths of mlcad\_d181\_lefttwo3rds and boom\_med\_pb are reduced by over 16\%.

\begin{table}[t]
\caption{The comparison between Ours w.o. RPTT and Ours. The ratios larger than 1 represent the quality degradation.}
\label{tab:exp_ablation_tree_enhance}
\resizebox{1.0\linewidth}{!}{
\begin{tabular}{l|ccc}
\hline
Benchmark & Runtime (s) & Wirelength & Score \\ \hline
logicnets\_jscl & 1.02 & 0.98 & 1.00 \\
boom\_med\_pb & 1.15 & 1.02 & 1.10 \\
vtr\_mcml & 1.46 & 1.06 & 1.30 \\
rosetta\_fd & 1.11 & 1.06 & 1.09 \\
corundum\_25g & 1.03 & 0.76 & 0.95 \\
finn\_radioml & 1.02 & 1.04 & 1.03 \\
vtr\_lu64peeng & 1.12 & 1.02 & 1.06 \\
corescore\_500 & 1.08 & 1.01 & 1.04 \\
corescore\_500\_pb & 1.11 & 1.08 & 1.10 \\
mlcad\_d181\_lefttwo3rds & 1.16 & 1.11 & 1.15 \\
koios\_dla\_like\_large & 1.14 & 1.04 & 1.12 \\
boom\_soc & 1.42 & 0.98 & 1.32 \\
ispd16\_example2 & 1.01 & 0.99 & 1.00 \\ \hline
Avg. Ratio & 1.14 & 1.01 & 1.10 \\ \hline
\end{tabular}
}
\end{table}

\subsection{Impact of The Number of Threads}
\begin{figure}[t]
    \centering
    \includegraphics[width=0.95\linewidth]{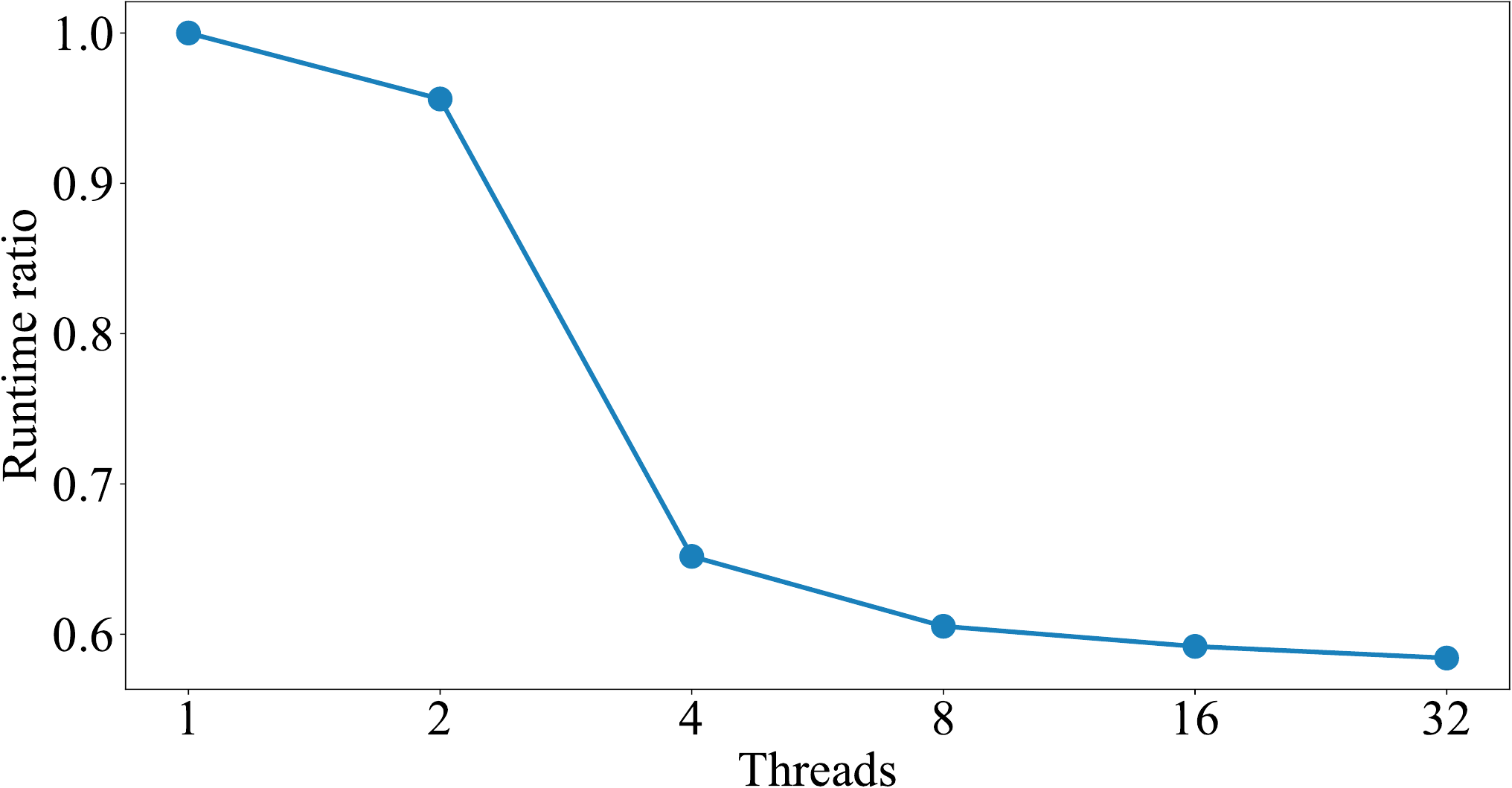}
    \caption{The average runtime ratio with different threads.}
    \label{fig:exp_thread}
\end{figure}

Furthermore, we also study the impact of the number of threads on our parallel router. As illustrated in Fig.~\ref{fig:exp_thread}, compared with the single thread, the runtime keeps reducing with the increase of thread number but will gradually converge at 32 threads.

%% file: sections/conclusion.tex
\section{Conclusion}
\label{sec:conclusion}
In this work, we propose a fast parallel routing approach for commercial FPGAs, which incorporates a recursive partitioning ternary tree and a hybrid updating strategy for congestion coefficients. Compared with the commercial tool, our proposed approach has achieved significant improvements in both runtime and wirelength.